\documentclass[10pt,acmtog, authorversion]{acmart}

\AtBeginDocument{%
  \providecommand\BibTeX{{%
    \normalfont B\kern-0.5em{\scshape i\kern-0.25em b}\kern-0.8em\TeX}}}

\usepackage{amsmath,amssymb,amsfonts}
\usepackage[ruled,vlined]{algorithm2e}
\usepackage{graphicx}
\graphicspath{ {./diagrams/} }
\usepackage{tikz}
\usepackage{pgfplots}
\usepackage{textcomp}
\usepackage{xcolor}
\usepackage{tcolorbox}
\usepackage{thmtools}
\usepackage{thm-restate}
\newtheorem{theorem}{Theorem}
\newtheorem{definition}{Definition}

\SetKwBlock{MachineParallel}{Machine Parallel For}{end}
\SetKwBlock{ParallelFor}{CPU Parallel For}{end}
\newcommand{\Lock}[1]{\textbf{Lock}(#1)}
\newcommand{\Unlock}[1]{\textbf{Unlock}(#1)}
\newcommand{\Insert}[2]{#1.\textbf{insert}(#2)}

\begin{document}

\title{Distributed Tera-Scale Similarity Search with MPI: Provably Efficient Similarity Search over billions without a Single Distance Computation}

\author{Nicholas Meisburger}
\email{ncm5@rice.edu}
\affiliation{
  \institution{Rice University}
  \streetaddress{6100 Main Street}
  \city{Houston}
  \state{Texas}
  \postcode{77005}
}

\author{Anshumali Shrivastava}
\email{anshumali@rice.edu}
\affiliation{
  \institution{Rice University}
  \streetaddress{6100 Main Street}
  \city{Houston}
  \state{Texas}
  \postcode{77005}
}

\begin{abstract}
We present SLASH (Sketched LocAlity Sensitive Hashing), an MPI (Message Passing Interface) based distributed system for approximate similarity search over terabyte scale datasets. SLASH provides a multi-node implementation of the popular LSH (locality sensitive hashing) algorithm, which is generally implemented on a single machine. We show how we can append the LSH algorithm with heavy hitters sketches to provably solve the (high) similarity search problem without a single distance computation. Overall, we mathematically show that, under  realistic data assumptions, we can identify the near-neighbor of a given query $q$ in sub-linear ($ \ll O(n)$) number of simple sketch aggregation operations only. To make such a system practical, we offer a novel design and sketching solution to reduce the inter-machine communication overheads exponentially. In a direct comparison on comparable hardware, SLASH is more than 10000x faster than the popular LSH package in PySpark. PySpark is a widely-adopted distributed implementation of the LSH algorithm for large datasets and is deployed in commercial platforms. In the end, we show how our system scale to Tera-scale Criteo dataset with more than 4 billion samples. SLASH can index this 2.3 terabyte data over 20 nodes in under an hour, with query times in a fraction of milliseconds. To the best of our knowledge, there is no open-source system that can index and perform a similarity search on Criteo with a commodity cluster. 
\end{abstract}




\maketitle

\section{Introduction}

Similarity search, or k-nearest-neighbor (k-NN) search, is one of the most frequently used operations in any large data processing system. The capability to quickly find data "similar" to a given query is the first prerequisite that enables innumerable applications such as information retrieval~\cite{das}, question answering~\cite{panchenko}, computational biology~\cite{parry}, etc. 
Formally, given a query object $q$, represented as feature vector $q\in \mathbb{R}^D$, the goal of similarity search is to find, from a collection $\mathcal{C}$ of $N$ data instances, an object $x$ (or set of objects) most similar to the given query. The notions of similarity are based on some popular Euclidean type measures such as cosine similarity~\cite{zahrotun} or Jaccard similarity~\cite{zahrotun}  

Similarity search on large data set is a well-studied application. There are innumerable algorithms and open-sourced system implementations, far more than many other topics in machine learning and big data mining. However, almost all of them are single machine algorithms that assume that the data can fit in the main memory of a single machine. Surprisingly, we found that all publicly available implementations fail to scale when the dataset under consideration is terabyte-sized or higher, which is our focus.

{\bf  Criteo Tera-Scale Dataset and Hardness of k-NN search:}  In 2013, Criteo released a terabyte size dataset referred to as the Terabyte Click Logs~\cite{libsvm}. For this dataset, even running a simple logistic regression is considered hard due to its enormous size~\cite{sterbenz}. 

The computational cost of k-NN search on this dataset is enormous. Criteo training data consist of around 4 billion samples. Each sample is a very sparse million-dimensional vector with approximately 40 nonzero values on an average.  The extensive, almost countless, categorical features are hashed~\cite{libsvm} to about a million dimensions for the feasibility of storing the data in a vector form.  A naive brute force k-NN search will require about 160 billion  (4 billion x 40 nonzeros) multiplication per query.  A well-known workaround is allowing approximations during the near-neighbor search.

{\bf FAISS and HNSW cannot work with sparse million dimensional vector:} Recently similarity search with 1 billion samples has gained good interest~\cite{johnson2019billion} with the development of FAISS (Facebook) and HNSW~\cite{malkov2018efficient} implementations. However, these implementations are for datasets with a few hundred dimensions that can fit on a single machine. The index size and indexing time associated with FAISS and HNSW make these systems far to expensive for a dataset of this size. Our finding echos what is observed in ~\cite{wang}. Here, the authors demonstrated, via rigorous experimental evaluation, that both FAISS and HNSW are orders of magnitude worse than naive LSH (Locality Sensitve Hashing) implementations for Criteo like very high-dimensional and sparse dataset. 

{\bf Curse of Dimensionality and Locality Sensitive Hashing (LSH):} It is well known that efficiency of approximate nearest-neighbor search (ANN) deteriorates as the dimensionality of the data increases. This phenomenon is known as the curse of dimensionality. KDTrees~\cite{zhou2008real} and related deterministic space partitioning techniques are particularly prone to the curse of dimensionality. Even with as a small as few hundred dimension's these techniques show worse performance compared to brute-force~\cite{weber1998quantitative}.   

In particular, randomized algorithms, based on LSH~\cite{indyk1998approximate}, are the only techniques that have shown promise on sparse high dimensional datasets with millions of dimensions. In ~\cite{wang}, through rigorous comparisons, it was shown that when the data vectors are sparse and high dimensionalal, applying LSH is orders of magnitude better than the best ANN packages, including FAISS and HNSW. The indexing time with  LSH is also quite appealing, as it is near-linear and embarrassingly parallel, which is rarely true with other alternatives. We do not see any hope of any algorithm, other than LSH, that can scale to 4 billion million-dimensional vectors occupying terabytes of memory. It is also worth noting that learning to hash~\cite{wang2017survey} methods cannot scale to this size of dataset either.

{\bf Open Source Distributed LSH Implementations:} Due to the significance of the similarity search problem, there are many distributed LSH packages available on GitHub. Almost all of them run on top of the Apache Spark framework and all of them essentially implement the same LSH algorithm (see Section~\ref{sec:lsh}) and its popular variants. Apache Spark framework poses several problems for efficient nearest neighbor search. Firstly, because spark runs on top of the Java Virtual Machine, there are computational overheads such as garbage collection and its associated operations that limit the performance on these computationally intensive tasks. More importantly however, is that the memory requirements of the system significantly reduce the ability to scale to larger datasets. By default, Apache Spark will only use 90\% of the available heap to avoid memory errors, and of that 90\%, only 60\% is made available for caching data \cite{grishchenko}. Effectively, we can only cache 540 megabytes of data for every gigabyte of heap space. For Criteo, these memory overheads are prohibitive. 

{\bf Apache Spark (PySpark) LSH cannot scale to Criteo on a Commodity Cluster:} Tera-scale similarity search (or ANN) is routinely solved in big data processing companies. The parallel nature of the problem allows the industry to solve this problem by deploying massive, expensive, distributed clusters. These systems do not scale well. The most popular tool for large scale distributed data analysis is PySpark. Recently a PySpark package for nearest neighbor search using LSH was released \cite{ni}. We found that this commercial-grade distributed LSH algorithm to be prohibitively slow for Criteo dataset. To get a sense of its scalability, we evaluated PySpark LSH on a much smaller scale.  We took a subset of webspam dataset \cite{libsvm}, having similar characteristics as Criteo, but much smaller. The PySpark LSH system is around 10000x slower than our proposed system SLASH on the same cluster size and number of CPUs. With PySpark, computing a simple set of 100 nearest neighbor queries on the webspam subset of around 10,000 vectors took on the order of 20 to 30 minutes. For comparison, the same task, on simple commodity system cluster using SLASH, took a fraction of a second (see Figure \ref{fig:spark_comp}). A batch of 10,000 queries on the full 350,000 webspam dataset using our system, could be done in a few seconds with SLASH. This result is indicative of the limitations of the PySpark LSH package, both in its inability to scale to large datasets and clusters, and its slow performance even on small queries. 

Unfortunately, PySpark is the best publicly available distributed implementations of LSH. PySpark is routinely deployed in several commercial big-data companies. For instance, it is deployed for online fraud detection in Uber ride transaction~\cite{ni}. 

{\bf Too many distance computations in the LSH Algorithm:} Given a query, the LSH algorithm (see Section~\ref{sec:lsh}) quickly prunes the potential candidate solution. However, due to high false-positive rates, the candidates need to be filtered using exact distance computations. This step is significantly substantial in practice. The size of the candidate set can be a significant fraction of the whole data, making the process almost a brute force search. ~\cite{wang,shrivastava2012fast} mitigates this issue using the frequency of occurrence as a cheap proxy for distances. However, the effect of this heuristic is theoretically not apparent.

{\bf Our Focus and Contributions:} Our goal is to build a distributed LSH system that can scale to Criteo or bigger dataset) on a commodity cluster.
Broadly this paper contributes both a novel  algorithm and a novel system implementation. Our work is an auspicious illustration of the power of smart algorithms combined with parallelism

{\bf 1. Algorithmic Contributions:} We create a novel heavy hitter sketches~\cite{cormode} from data streams literature with classical LSH algorithm.
This integration creates a provably novel algorithm that can solve high-similarity search, over massive datasets, efficiently in sub-linear cost.  We show that while creating the LSH data structure, we can replace LSH buckets with small fixed-sized heavy-hitter sketches. Effectively, the similarity search query boils down to a sub-linear number of fixed-sized sketch aggregation for approximating the most frequent elements. We show that these approximate frequent elements retrained by the algorithm are guaranteed to be the nearest-neighbors of the query. Our algorithm is the first sub-linear similarity search algorithm that does not require any distance computations to the best of our knowledge. 
We provide precise asymptotic mathematical quantification of the number of operations needed to find the nearest-neighbors with high probability. 

{\bf 2. Scalable MPI Implementation:} To avoid the Apache Spark framework's unnecessary overhead, we chose MPI (Message Passing Interface), which is popular in the HPC (High Performance Computing) community. SLASH is an MPI implementation of LSH, which could be of independent interest in itself. In addition to novel design choices, we leverage recently proposed TopKapi sketches to replace hash buckets. This choice reduced the memory footprint of the buckets and also eliminates load variability due to variable bucket sizes.  In addition, the unique mergeable property of the sketches allows us to perform bucket aggregation much more efficiently, leveraging the logarithmic scaling of reduction style communication. Through rigorous experiments, we demonstrate a remarkable jump in the capabilities to do distributed similarity search. Our system, with mere 20 nodes (12 cores each), can  \textbf{index\ 4 billion samples} in about \textbf{40 minutes}, with \textbf{query time in a fraction of milliseconds}.   

\section{Background}

\subsection{Locality Sensitive Hashing}
Locality sensitive hashing (LSH) is a hashing scheme with the important property that the probability of a hash collision for vectors $x,y \in \mathbb{R}^d$ is inversely related to the distance between $x$ and $y$. Given a hash family $\mathcal{H}:= \{h: x \to \mathbb{R}\}$. We say that $\mathcal{H}$ is a locality sensitive hash function if, given a distance function $D$ (e.g. cosine or Jaccard distance), and $x_1, x_2, y_1, y_2 \in \mathbb{R}^d$ ~\cite{har-peled}.
$P(h(x_1) = h(y_1)) \geq P(h(x_2) = h(y_2)) \Leftrightarrow \ \ D(x_1, y_1) \leq D(x_2, y_2)$

\subsection{MinHash and Densified MinHash}
MinHash is a well known and established locality sensitive hashing scheme. We can define the MinHash of a vector as follows \cite{shrivastava}: Given a binary vector $x = \{0,1\}^d$, and a random permutation, $\pi$, of the elements of $x$, the MinHash of $x$ is: $h(x) = \min_{x_i = 1} \pi(i)$

Furthermore, if $x,y \in \{0,1\}^d$ then for the MinHash function, $h$, then the probability of a hash collision between $x$ and $y$ is the Jaccard index of the vectors \cite{shrivastava}: $P(h(x) = h(y)) = \frac{x \cdot y}{|x| + |y| + x \cdot y}.$ Recently it was shown, both theoretically and empirically, that for sparse data min-hash is superior to random projection based hash functions~\cite{shrivastava2014defense}. 


One major drawback of  standard MinHash is the cost of computing several hashes for every vector in the dataset during processing, and every query vector during querying. For a similarity search system using MinHash the cost of computing multiple hashes of the query itself, requires hundreds or thousands of passes over the data vector, which is known to be the bottleneck step while querying~\cite{shrivastava2017optimal}. A solution to this problem is \textbf{Densified One Permutaion Hashing (DOPH)} which allows you to compute thousands of hashes in one pass over the given vector~\cite{shrivastava}, essentially reducing all the hashing time to data reading time.

In order to compute $N$ hashes of a given vector using DOPH, simply partition the range of the hash function into $K$ buckets, then hash each of the non-zero indices of the vector and select the minimum hash that falls within each bucket as each of the hashes. If a bucket is empty, densify the hashes by selecting the hash of the closest non empty bucket on either the left or right (choosen randomly).

\subsection{The LSH Algorithm}
\label{sec:lsh}
For similarity search, LSH values of a data vector is used for indexing (or keys). Two vectors with same keys are very likely to be similar. We discuss the common formulation of the $(K,\ L)$ parametrized LSH algorithm~\cite{andoni} that is structured as follows. Given values of two parameters $K$ and $L$, create $L$ hash tables, each with a hash function $H_i$ for $i \in [1,2,...,L]$. Each of these hash functions, $H_i$ is constructed using $K$ LSH functions, $H_i = \{h_{i_1}, h_{i_2},...,h_{i_K}\}$ such that foreach vector $x\in \mathbb{R}^d$ $H_i: x \to \mathbb{R}^K$ ~\cite{andoni}. 

There are two distinct phases for processing a dataset $D$:
\begin{enumerate}
     \item \textbf{Pre-processing Phase}: For vector $x \in D$, for hash table $i \in [1,2,...,L]$ compute $H_i(x)$ which will be a vector in $\mathbb{R}^K$. Use this vector to index $x$ and store an identifier for $x$ at this hash bucket in table $i$.
    \item \textbf{Query Phase}: The query process is usually online. Here, we have a query vector $q$. It is composed of two parts:
    \begin{enumerate}
        \item \textbf{Bucket Aggregation}: For a given query vector $q$ compute $[H_1(q), H_2(q),...,H_L(q)]$ and construct the candidate set by taking the union of the hash buckets corresponding to these locations in each of their respective hash tables.
        \item \textbf{Filtering and K-NN Selection}: The candidate set, generated in aggregation step, is searched for the closest candidate to the query. This step can be exact or approximate.
    \end{enumerate}
\end{enumerate}

{\bf Cheap Frequency based Filtering and K-NN:}  Because the probability of a hash collision of vectors, $x,y \in \mathbb{R}^d$ is greater if $x$ and $y$ have a smaller distance between them, the number of collisions between $x$ and $y$ across all $L$ hash tables is larger if $x$ and $y$ are closer. Thus we can replace an exact distance based ranking of selected candidates with a simple frequency count of number of occurrences of each candidate vectors in the $L$  buckets hashed by the query. The idea is has been used at several LSH implementations in the past~\cite{shrivastava2012fast,wang}.





\subsection{TopKAPI}

We will use TopKAPI sketch for estimating the top-k frequent elements. TopKAPI~\cite{mandal} is a simple combination of frequent items~\cite{karp2003simple}  and count sketch~\cite{cormode} based approach which combines the best of the two worlds. It results in a sketching algorithm which is ideal for parallelism and is orders of magnitude faster than best existing implementing of either. We chose TopKAPI because it can exploit both local (multi-core) and distributed parallelism ideally suited for MPI implementation. Due to space constraints, we give a high lever overview of this prior background work in appendix. For details on why it works and comparisons with existing sketching approaches see~\cite{mandal}

\section{Our System: SLASH}

Our system is a distributed implementation of the LSH algorithm. We call our system SLASH. We first describe the design of SLASH. Later we discuss the algorithmic piece which involves sketching and aggregation. Later in section~\ref{sec:correctness}, we show the correctness of the algorithm and analyse its computational cost. 

\paragraph{Design: Distribute Data and Hash Tables Across Machines:} Our design is simple and ensures no inter-node communication, except a one time sharing of the same random seed for the algorithm, during index creation. No communication between machines ensures best scalability, which is critical and otherwise impossible with sophisticated nearest-neighbor algorithms other than LSH. 

\begin{figure*}
    \centering
    \includegraphics[scale=0.4]{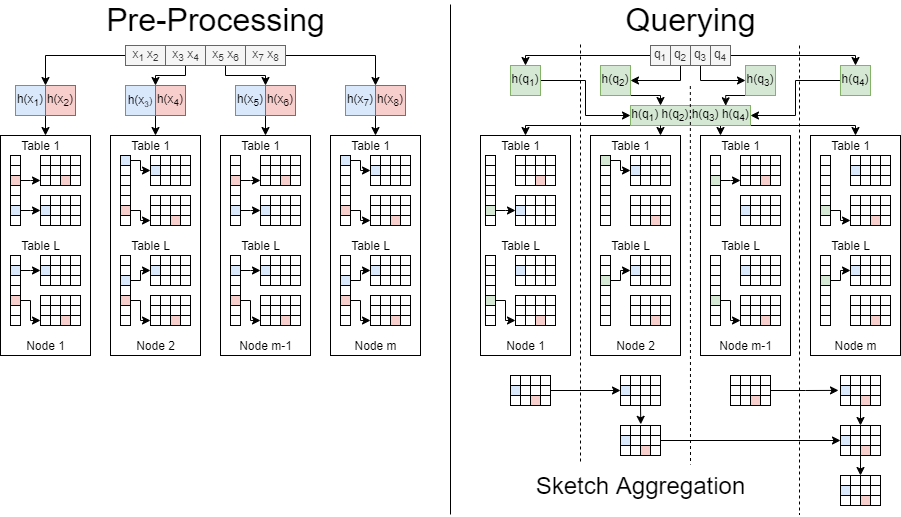}
    \caption{Diagram of the SLASH algorithm. Data is partitioned across each node and indexed during preprocessing. During queries, a batch of queries is hashed in parallel, these hashes are distributed and the corresponding sketches are retrieved and merged.}
    \label{fig:diagram}
\end{figure*}

Our design is illustrated in Figure~\ref{fig:diagram}. Assuming $m$ computing nodes. We have $m$ data partitions, call it ${D_1, \ D_2, \ D_3, ..., \ D_m}$, one for each node. Each computing node, i, creates $(K,\ L)$ parameterized LSH data structures for each of the partition $D_i$ of the dataset. To ensure we have the same LSH function, we simply start with the same random seed, a unique advantage of randomized algorithms. As a result of these consistent hash functions, there is no need to re-hash vectors on each node at query time. 

By sharing the random seed, we ensure same LSH hash functions across the nodes. As a result, our design is equivalent to having one  $(K, \ L)$ parameterized LSH data  structure for the complete data $D = \bigcup_{i =1}^m D_i$, and each bucket of the hash tables is disjointly partitioned across computing nodes. Furthermore, the partitioning is done in a way that is  consistent with the local data of the computing node.

{\bf Algorithmic Challenges to LSH} Here we mention two algorithmic challenges that need to be addressed for the LSH system to make full use of parallelism.

{\bf Avoid Skewed buckets: } LSH buckets are known to be highly skewed. Especially, heavy buckets slow down the performance significantly by creating variable workloads during aggregation. This is a well-known problem. To mitigate this variability, authors in~\cite{wang} chose to sub-sample the buckets, using online reservoir sampling~\cite{vitter},  to a fixed size if they are heavy. 

First of all, sub-sampling to create fixed sized buckets, by randomly throwing elements from large buckets, is likely to lose accuracy. It was argued in~\cite{wang} that sub-sampling retains the LSH property, and the loss due to sampling can be compensated by having a larger number of tables. However, a larger number of tables (or large values of $L$) causes memory and computational overhead during pre-processing and querying. At the same time, the buckets mustn't be skewed for maximal utilization of parallelism during aggregation. Skewed buckets have a very high chance of having stragglers, where most processes are waiting for one process to finish.

\textbf{Avoid Expensive Linear Communications:} Even with a cheap frequency-based re-ranking, which involves taking the weighted union of the L (the number of hash tables) buckets from each node followed by sorting, the communication and computation overhead across each node is significant. It should be noted that this procedure needs to be performed for each query and directly affects the query cost. In ~\cite{wang}, the process was tailored for multi-core machines where all the threads work on a shared main memory where all the hash tables reside. There, communication between different CPU processes did not create a significant performance bottleneck.  However, in a distributed environment hash tables live on different nodes, as they won't fit on a single machine. Here, communication is the main bottleneck.  

\subsection{Our Solution: Topkapi Sketches instead of Buckets}
\label{sec:topkapi}

It turns out that there is one principled solution for all the above problems.  We replace the buckets by a recently proposed top-k frequency counting sketch Topkapi~\cite{mandal}. 

\textbf{Why Topkapi and not other sketches?} Topkapi is a distributed variant of the famous count-min sketch~\cite{cormode} and is a streaming algorithm for reporting the identity of the approximate top-k most frequent element in any stream. Effectively, Topkapi sketch consists of small fixed-sized arrays than can incrementally ingest any stream. It supports two operations: 1) add an element to the stream and 2) report the identity of top-k most frequent elements. The size of Topkapi is significantly smaller than the buckets and are fixed size. The query and insertion time is independent of the size of the sketch. However, Topkapi has a unique property of \emph{mergeablity} that has unique advantages in a distributed setting. Mergeablity implies that sketches on two disjoint partitions can be merged to obtain a sketch for the union of the two partitions. It was noted in~\cite{mandal} that popular sketches, including count-min sketch, are not mergeable as the top-k of the union cannot be obtained from the top-k of the individual sets. Topkapi combines the advantages of count-min sketch with lossy counting that leads to an efficient, parallelizable, and mergeable sketch. Please see~\cite{mandal} for more details and performance comparisons in distributed environments.  

\textbf{Bucket size reduction with perfectly balanced aggregation (no skew):} Topkapi can be updated in a streaming manner. We can replace the buckets with a fixed-sized Topkapi. Instead of adding the element in the bucket, we increment the sketch. Additionally, merging and top-k querying is constant time operations, so the overhead is minimal. Most importantly, the size of the sketch does not grow with addition. Thus, no matter how many elements are inserted in the sketch, its size remains the same. The cost of all operations on the sketch also remains the same. With Topkapi, there is a perfect balance of computations while updating as well as aggregating buckets from different hash tables, despite the skew in the number of elements. Thus, unlike~\cite{wang}, there is no need for sub-sampling the buckets, where the chance of losing good candidates increases. A small fixed-sized Topkapi is sufficient for efficient and accurate retrieval. The size of the sketch only depends on the number of \emph{good neighbors} and not $n$.

\begin{algorithm}
\caption{Preprocessing with SLASH}
\label{alg:preprocess}
\nl \MachineParallel($n \in \{1...m\}$){
    \nl \ParallelFor($x \in D_n$) {
        \nl \For{$i \in \{1....L\}$}{
            \nl $h \gets H_i(x)$\\
            \nl \Lock{$T_i(h)$}\  \textit{// The Topkapi Sketch} \\
            \nl \Insert{$T_i(h)$}{$x$}\ \ \ \  \textit{// Sketch Insert} \\
            \nl \Unlock{$T_i(h)$}\\
        }
    }
}
\end{algorithm}
\textbf{Allreduce due to Mergeablity:} The operation of bucket aggregation requires finding top-k most frequent elements in all the accumulated buckets from all the nodes. First of all, the small size of sketch reduces the memory footprint of communication.  Furthermore, due to mergeablity property of Topkapi sketches, this operation reduces to an Allreduce style sketch merging of Topkapi. The Allreduce operation only requires logarithmic wait time in communication to complete instead of linear time.

\section{Overall Algorithm and Implementation Details} 

We provide details of pre-processing and querying steps with SLASH. 

During query, the aim of LSH (or any indexing) is to narrow down the candidate set to a small enough set of hash buckets. After this reduction, we use brute force distance computation and sorting to obtain nearest neighbors from this small set. One solution to intelligently avoid computing distances is leveraging the observation that the probability of hash collisions in a locality sensitive hash function is directly proportional to the similarity of the vectors ~\cite{wang}. As a result, for a candidate, we can simply use its frequency of occurrence in the selected $L$ buckets as a crude measure for ranking (as explained in Section~\ref{sec:lsh}). This is a neat observation that has been used in the past to reduce the aggregation time~\cite{shrivastava2012fast,wang}. It avoids the costly distance computations. We will retain this same ranking algorithm. We will also retain the use of DOPH as the hash function, as it is the fastest and most accurate hash function for high-dimentional and sparse dataset of interest. 

However, there are two major shortcomings of this approach that become critical when it comes to parallelism and distributed workloads. It is know that LSH buckets vary drastically in size. Some buckets are quite heavy while most others are near-empty. Secondly, distributed frequency counting on large buckets can quickly become expensive by requiring extensive communication.

Our system solves this problem by replacing each reservoir with a TopKAPI sketch. Inserting elements into these sketches requires only a couple of simple hash computations, and so it adds very little computation overhead while offering significant benefits in the query process. We can break down our algorithm into 2 phases: indexing and querying. 

\textbf{Initialization:} We have $m$ computing nodes and $m$ disjoint partition of the data $D$, given by $\{D_1, \ D_2, ..., \ D_m\}$. We initialize $L \times K$ different hash functions $h_{i_j}$, where $i \in \{1, \ 2,..,\ L\}$ and $j \in \{1, \ 2, ..., \ K \}$, by initializing $L \times K$ random seeds. Effectively, all the computing nodes share these seeds, i.e., they all have the same hash functions. On each computing node, initialize $L$ hash tables, each with a hash function $H_i$ for $i \in [1,2,...,L]$. Every address $a$, in the hash table $T$ is associated with a Topkapi sketches $T(a)$, instead of usual hash buckets. Each of the hash functions, $H_i$, is constructed using $K$ Locality Sensitive hash functions, $H_i = \{h_{i_1}, h_{i_2},...,h_{i_K}\}$ such that foreach vector $x\in \mathbb{R}^d$ $H_i: x \to \mathbb{R}^K$ ~\cite{andoni}. 

\textbf{Pre-processing Phase}: 
On each machine $k$, we process its corresponding data partition $D_k$ independently in parallel. For each vector $x \in D_k$, foreach hash table $i \in [1,2,...,L]$ compute $H_i(x)$ which will be a vector in $\mathbb{R}^K$. Use this vector to index $x$, and insert $x$ into the Topkapi sketch $T_i[H_i(x)]$, associated with the address $H_i(x)$ in table $T_i$. The procedure is summarized in Algorithm \ref{alg:preprocess}.

The key benefit of this approach is that the size of the model is reduced in each node as each node no longer needs to be able to index and store the full data set. This model parallelism means that each node added to the cluster either decreases the memory requirement in each node by further dividing the data set, or allows for an even larger model by increasing the overall memory capacity of the system.

We can also obtain a speedup in this phase by partitioning the dataset into separate files. This allows us to access data from each file in parallel without concerns with synchronization or separate machines interfering with each other. In distributed MPI jobs, having multiple machines access the same file simultaneously can hurt the performance of parallel file systems. One solution to this is to use the MPI defined file IO library which solves the problem by synchronizing accesses between machines. We observed that since different machines were always accessing disjoint data we could partition the datasets to allow for maximum parallelism without any synchronization. This, combined with maintaining a large internal buffer on each node to reduce reads to disk allowed us to improve the indexing time of our system from hours to minutes.  

\textbf{Querying:}
Because our LSH model is distributed, in order to query the nearest neighbors of a given vector, we must aggregate the sketches on a local and global level. Once the hash indices of a query are computed, each node locates the sketches stored at corresponding index and merge all of them into a final sketch. The final sketch is queried for top-k. The process is summarized in Algorithm~\ref{alg:query}.

\begin{algorithm}
\caption{Querying with SLASH}
\label{alg:query}
\KwIn{A batch of queries, $Q = \{Q_1, Q_1, ..., Q_m\}$}
\nl $I \gets \{I_1, I_2, ... I_m\}$\ \ \ \ \textit{// Stores the hash indices per partition}\\
\nl \MachineParallel($n \in \{1...m\}$){
    \nl \ParallelFor($q \in Q_n$) {
        $I_{n_q} \gets \bigcup_{i \in \{1...L\}} H_i(q)$\ \ \ \ \textit{// Compute hashes}\\
    }
}
\nl {\bf MPI\_Allgather}$(I)$\ \ \ \ \textit{// Tranfer hashes}\\
\nl $S \gets$ array of empty sketches in each machine\\
\nl \MachineParallel($n \in \{1...m\}$) {
    \nl \ParallelFor ($q \in I$) {
        \For{$i \in \{1...L\}$}{
            \nl $S_q \gets${\bf Merge}($S_q$, $T_i(I_{q_i})$)\ \ \ \ \textit{// Sketch Merge}\\
        }
    }
}
\nl $A \gets \{1,2,...m\}$\ \ \ \ \textit{// Represents active machines}\\
\nl \MachineParallel($n \in \{1...m\}$) {
    \nl \For{$\{1...\log(m)\}$}{
        \nl $r \gets A$.{\bf indexOf}($n$)\ \ \ \ \textit{// Check active machine}\\
        \nl \If{$r \mod 2 = 0$}{
            \nl {\bf MPI\_Send}($S$) {\bf To } $A[r-1]$\ \ \textit{// To previous active}\\ 
            \nl $A$.{\bf Remove}($r$)\ \ \ \ \textit{// Machine Inactive}\\
        }
        \nl \ElseIf{$r \mod 2 = 1$}{
            \nl $S' \gets$ {\bf MPI\_Recv}() {\bf From } $A[r+1]$\   \textit{//  next machine}\\
            \nl $S \gets${\bf Merge}($S$, $S'$) \ \ \ \ \textit{// Pairwise merge }\\
        }
    }
}
\nl {\bf Machine 1 Returns} $S$\\
\end{algorithm}

Locally merging sketches uses local parallelism, which does not have much overhead. However, merging sketches across the cluster is a more complex. Using an MPI\_Gather call to copy all of the sketches to a single node and then merging them into a final sketch may seem like a natural idea. However, MPI\_Gather only uses parallelism for the message passing, the computational work for merging sketches and is done on a single machine, which creates stragglers. We implemented a custom MPI reduction pattern that uses MPI primatives to integrate the sketch merging with the MPI communication. At each time step a given node would copy its current sketch to the preceding node where it would be merged with that node's sketch. These communications were synchronized so that the message passing and merging happened in parallel in different parts of the cluster (see Figure ~\ref{fig:diagram}) creating a tree pattern reduction, similar to Allreduce. Because sketch merging accounted for the majority of the time in this phase of the query, we found that our query times went down logarithmically when we uses this tree based aggregation method. 

\section{Analysis of Algorithm: Correctness and Computation Complexity}
\label{sec:correctness}

\subsection{LSH Preliminaries}
Our primary notations and definitions of $(c,r)$-approximate nearest neighbor are covered in full rigor Section 1.2 in~\cite{har-peled}. We will simplify them and present them for the sake of completeness. We will use $D(x,y)$ to denote the distance between $x$ and $y$. The distance measure can be any metric.

\begin{definition}~\cite{har-peled}
The {\bf $(c,r)$-approximate near-neighbor problem (or $(c,r)-NN)$} with failure probability $f$ is to construct a data structure over the dataset supporting the following query: given any fixed query point $q$, if there exist $x$ in the dataset such that $D(q,x) \le r$, then report some $x'$ such that $D(q,x') \le cr$, with probability $1-f$. 
\end{definition}

We always have $r \ge 0$ and $c \ge 1$. We skip the failure probability $f$ in this or other definitions involving it if it is equal to some absolute constant from $(0,1)$. We can always boost this to any small constant by union bound and paying an extra logarithmic cost. 

\begin{definition}~\cite{har-peled}
We call an LSH hash function $(r,cr,p_1,p_2)$-sensitive if for any $x_1$ and $x_2$, we have two conditions:
\begin{enumerate}
    \item if $D(x_1,x_2) \le r$ then $Pr(h(x_1) = h(x_2)) \le p_1 $
    \item if $D(x_1,x_2) \ge cr$ then $Pr(h(x_1) = h(x_2)) \ge p_2 $
\end{enumerate}
\end{definition}

Our algorithm replaces LSH buckets with a principled sketching algorithm for approximate top-k aggregation. It turns out that under minor assumptions, we can show that the running time of our Algorithm~\ref{alg:query} is sub-linear. For the sake of brevity, we only state the main result. The detailed proof is presented in the appendix. 

We will use the standard notions of $(c,r)$-approximate near-neighbor instance and the standard definition of $(r,cr,p_1,p_2)$-sensitive LSH. See~\cite{har-peled} for a good reference. Since we are using an approximate heavy hitter algorithm for cheap re-ranking, we need one extra condition to guard the signal-to-noise (SNR) ratio for identifiability. 

\begin{definition}\vspace{-0.1in}
We call a $(c,r)$-approximate near-neighbor instance k-bounded if the number of samples in the dataset within the distance of $cr$ from the query $q$ `is less than or equal to $k$.
\end{definition}

The $k$-bounded condition is very reasonable as we do not expect the significantly large number of samples to be close to the query. Otherwise the near-neighbor problem is anyway hopelessly hard. See~\cite{rubinstein2018hardness} for more details. Finally, we have our main result. The result establishes both correctness (success with  high probability) and running time of Algorithm~\ref{alg:query}. 

\begin{theorem} {\bf [Main Result]}
Assuming independently and identically distributed data samples (i.i.d.) of size $n$ with lar
ge enough value of $n$. Given a $k$-bounded $(c,r)$-approximate near-neighbor instance and a corresponding $(r,cr,p_1,p_2)$-sensitive LSH function, with $\frac{\log{p_1}}{\log{p_2}}  < 0.5$ (Strong LSH Instance), Algorithm~\ref{alg:query} with $L = n^{\rho}$ and $K= \frac{\log{n}}{\log{\frac{1}{p_2}} - \log{\frac{1}{p_1}}}$, where $\rho = \frac{\log{\frac{1}{p_1}}}{\log{\frac{1}{p_2}} - \log{\frac{1}{p_1}}} < 1$ identifies the true $r$-near-neighbors of any given query $q$ (point with $D(x,q) \le r$) with high probability. Essentially, the total query cost is precisely the cost of $O(n^\rho)$ heavy-hitters sketch aggregation of size $O(k)$. Unlike the well-known LSH procedure, there is no distance computations involved nor any post-filtering. 
\end{theorem}
{\bf Proof:} See Appendix. 

{\bf Asymptotic Running time:} Note Algorithm~\ref{alg:query} requires probing $L$ buckets, $KL$ hash functions evaluations, and finally $L$ bucket aggregation of fixed size $k$. Also, the bucket aggregation is logarithmic time $O(\log{L})$ and not $O(L)$ time because of mergeability.  Given the theorem, they all are asymptotically less than $O(n)$.     

{\bf Zero Distance Computations:} LSH algorithm requires sub-linear $O(L)$ distance computations. Our algorithm need $L$ bucket aggregation of fixed size $k$. This bucket aggregation costs only logarithmic time $O(\log{L})$ and not $O(L)$ time because of mergeability (see Section~\ref{sec:topkapi}) for details. As a result, we save significantly compared to existing implementations. 

{\bf Comment on The condition  $\frac{\log{\frac{1}{p_1}}}{\log{\frac{1}{p_2}}}  < 0.5$.} Small $\rho$  implies that this idea is only good for strong LSH, which are efficient in practice. LSH are commonly deployed only in such strong cases. Weak LSH with $\rho > 0.5$ are not practical even for medium scale, for tera-scale it is prohibitive.

\begin{table*}
\centering
    \begin{tabular}{|c|c|c|c|}
        \hline
         Name & \# Data Points & \# Features & Random Cosine Similarity  \\
         \hline
         Criteo & 4,195,197,692 / 178,274,637 & 1,000,000 & 0.12 \\
         \hline
         KDD12 & 149,639,105 & 54,686,452 & 0.15 \\ 
         \hline
         Webspam & 350,000 & 16,609,143 & 0.33 \\
         \hline
    \end{tabular}    
    \caption{Datasets Statistics and Mean Similarity}
    \label{tab:data}
    \vspace{-0.3in}
\end{table*}
\textbf{Corollary for Near-Duplicate Detection (De-Duplication):} For near-duplicate detection~\cite{xiao2011efficient,berrocal2016exploring}, $p_1$ is close to 1. With $p_1$ close to 1 the ,query time degenerates to logarithmic time!

\textbf{I.I.D. Assumption:} It is a very standard assumption used in machine learning, where every data is assumed to be i.i.d. sampled from some unknown distribution.

\section{Evaluations and Results}

{\bf Datasets Used:} In addition to tera-scale Criteo, we tested our system on the KDD12 and Webspam datasets. KDD12 is a clickthrough dataset and Webspam is dataset containing spam and legitiment emails.
\cite{libsvm}. The statistics of different data-sets is summarized in Table~\ref{tab:data}

\begin{figure*}
  \centering
\begin{tikzpicture}[scale=0.85]
\begin{axis}[
    title={Query Similarity},
    xlabel={Nodes},
    ylabel={Similarity},
    xmin=2, xmax=9,
    ymin=0.5, ymax=0.8,
    xtick={2,4,6,8},
    ytick={0.5, 0.6, 0.7, 0.8},
    legend pos=north west,
    ymajorgrids=true,
    grid style=dashed,
]
 
\addplot[
    color=red,
    mark=square,
    ]
    coordinates {
    (3, 0.64377)(4, 0.64377)(5, 0.64377)(6, 0.64377)(7, 0.64377)(8, 0.64377)
    };
    \addlegendentry{Spark}

\addplot[
    color=blue,
    mark=square,
    ]
    coordinates {
    (3, 0.627)(5, 0.627)(6, 0.626)(7, 0.627)(8, 0.625)
    };
    \addlegendentry{D-FLASH}
 
\end{axis}
\end{tikzpicture}
\begin{tikzpicture}[scale=0.85]
\begin{semilogyaxis}[
    title={Query Time},
    xlabel={Nodes},
    ylabel={Time (Seconds)},
    xmin=2, xmax=9,
    ymin=0.001, ymax=100000,
    xtick={2,4,6,8},
    ytick={0.001, 0.01, 0.1, 1, 10, 100, 1000, 10000, 100000},
    legend pos=north east,
    ymajorgrids=true,
    grid style=dashed,
]
 
\addplot[
    color=red,
    mark=square,
    ]
    coordinates {
    (3, 936.74585)(4, 908.79681)(5, 716.19876)(6, 473.65510)(7, 477.94534)(8, 477.13810)
    };
    \addlegendentry{Spark}

\addplot[
    color=blue,
    mark=square,
    ]
    coordinates {
    (3, 0.0238)(5, 0.0339)(6, 0.0327)(7, 0.0233)(8, 0.0305)
    };
    \addlegendentry{D-FLASH}
\end{semilogyaxis}
\end{tikzpicture}
\caption{Comparisons to PySpark LSH package. PySpark is around 10000x slower for the same retrieval quality}
\label{fig:spark_comp}
\end{figure*}
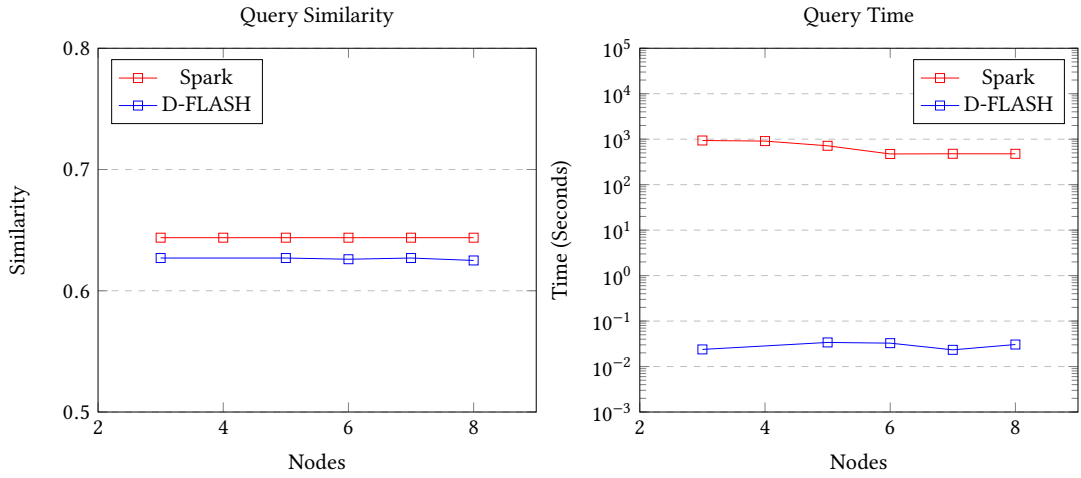

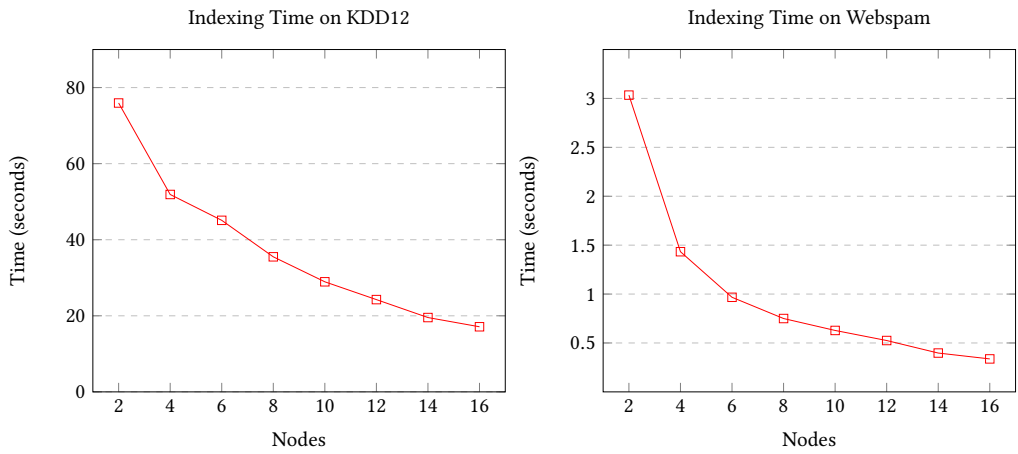
\begin{figure*}
\centering
\begin{tikzpicture}[scale=0.8]
\begin{axis}[
    title={Indexing Time on KDD12},
    xlabel={Nodes},
    ylabel={Time (seconds)},
    xmin=1, xmax=17,
    ymin=0, ymax=90,
    xtick={2,4,6,8,10,12,14,16},
    ytick={0,20,40,60,80},
    legend pos=north west,
    ymajorgrids=true,
    grid style=dashed,
]
 
\addplot[
    color=red,
    mark=square,
    ]
    coordinates {
    (2, 75.94299999999998)(4, 51.89500000000001)(6, 45.09800000000001)(8, 35.50410000000001)(10, 28.937399999999997)(12, 24.258199999999995)(14, 19.542)(16, 17.115999999999985)
    };
 
\end{axis}
\end{tikzpicture}
\begin{tikzpicture}[scale=0.8]
\begin{axis}[
    title={Indexing Time on Webspam},
    xlabel={Nodes},
    ylabel={Time (seconds)},
    xmin=1, xmax=17,
    ymin=0, ymax=3.5,
    xtick={2,4,6,8,10,12,14,16},
    ytick={0.5,1,1.5,2,2.5,3},
    legend pos=north west,
    ymajorgrids=true,
    grid style=dashed,
]
 
\addplot[
    color=red,
    mark=square,
    ]
    coordinates {
    (2, 3.033999999999992)(4, 1.4329999999999927)(6, 0.9670000000000414)(8, 0.7498999999999967)(10, 0.6280000000000143)(12, 0.5250000000000057)(14, 0.39700000000000557)(16, 0.3369999999999891)
    };
 
\end{axis}
\end{tikzpicture}
\caption{Indexing time Scaling with Computing nodes.}
\label{fig:indexing}
\end{figure*}
{\bf Quality Metric:} Following~\cite{wang}, we report the average cosine similarity~\cite{zahrotun} of the top-k reported results concerning the query datapoint as the quality metric, referred to as {\bf S@k}. We use the value of $k=1$ and $k=64$. Clearly, higher value of S@k indicate good quality. For LSH, recall is not a good measure because LSH can distinguish between two very close by neighbors. For many similarity search problem high value of S@k is sufficient~\cite{li2002clustering}.\\
{\bf Performance Metric:} We use the end-to-end wall clock indexing time and  querying time.\\
{\bf Hash Function:} We use MinHash~\cite{broder1997resemblance,indyk2001small} for all. SLASH implements the faster densified variants~\cite{shrivastava2017optimal}\\
{\bf Values of $K$ and $L$:} We keep the value of $K$ fixed at 4 for all the three datasets. For $L$ on webspam, KDD, and Criteo we use the values $24$, $16$ and $32$ respectively.\\
{\bf CPUs and Hardware:} Most SLASH experiments are on a commodity cluster of up to 20 nodes. This is a shared cluster with several users, so for the ease of scheduling, we limit the number of CPUs.  We used 4 CPUs for KDD12 and webspam, including PySpark.  For Criteo we used 12 cores. PySpark  was done on dedicated AWS with comparable computing CPU cores and nodes.  

\subsection{Comparisons with PySpark}

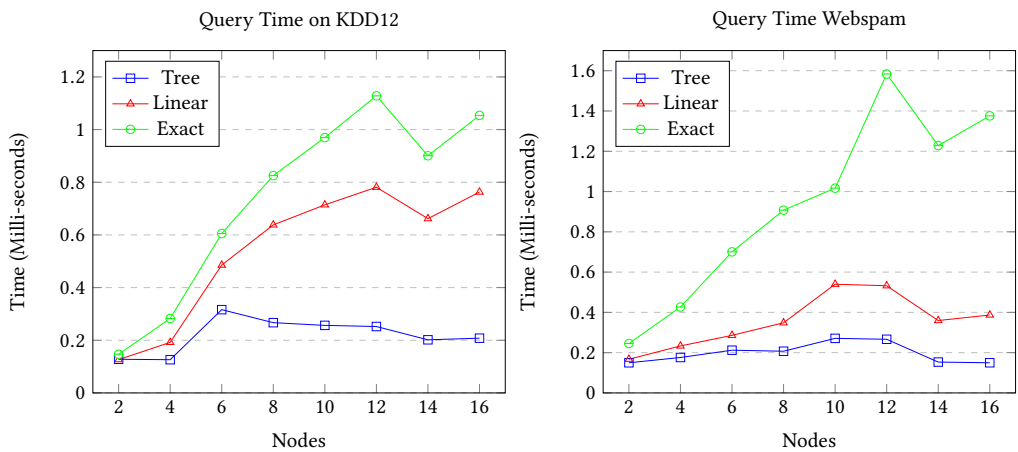
\begin{figure*}
\centering
\begin{tikzpicture}[scale=0.8]
\begin{axis}[
    title={Query Time on KDD12},
    xlabel={Nodes},
    ylabel={Time (Milli-seconds)},
    xmin=1, xmax=17,
    ymin=0, ymax=1.3,
    xtick={2,4,6,8,10,12,14,16},
    ytick={0,0.2,0.4,0.6,0.8,1.0,1.2},
    legend pos=north west,
    ymajorgrids=true,
    grid style=dashed,
]
 
\addplot[
    color=blue,
    mark=square,
    ]
    coordinates {
    (2, 0.12784295)(4, 0.12618723)(6, 0.316205635)(8, 0.26679842000000002)(10, 0.256528033)(12, 0.252178856)(14, 0.20167824)(16, 0.20790431999999996)
    };
    \addlegendentry{Tree}
 
 \addplot[
    color=red,
    mark=triangle,
    ]
    coordinates {
    (2, 0.12645795)(4, 0.19189823)(6, 0.485453635)(8, 0.63800342)(10, 0.714203033)(12, 0.781062856)(14, 0.66177924)(16, 0.76274032)
    };
    \addlegendentry{Linear}
    
    \addplot[
    color=green,
    mark=halfcircle,
    ]
    coordinates {
    (2, 0.14698195)(4, 0.28266023)(6, 0.605348635)(8, 0.82530742)(10, 0.969646033)(12, 1.1280928560000001)(14, 0.9005662399999999)(16, 1.0538613199999999)
    };
    \addlegendentry{Exact}
\end{axis}
\end{tikzpicture}
\begin{tikzpicture}[scale=0.8]
\begin{axis}[
    title={Query Time Webspam},
    xlabel={Nodes},
    ylabel={Time (Milli-seconds)},
    xmin=1, xmax=17,
    ymin=0, ymax=1.7,
    xtick={2,4,6,8,10,12,14,16},
    ytick={0,0.2,0.4,0.6,0.8,1.0,1.2,1.4,1.6},
    legend pos=north west,
    ymajorgrids=true,
    grid style=dashed,
]
 
\addplot[
    color=blue,
    mark=square,
    ]
    coordinates {
    (2, 0.14972443)(4, 0.17594049)(6, 0.21216211)(8, 0.20709243)(10, 0.27100074)(12, 0.26641501)(14, 0.15323362)(16, 0.14980563)
    };
    \addlegendentry{Tree}
 
 \addplot[
    color=red,
    mark=triangle,
    ]
    coordinates {
    (2, 0.16728643)(4, 0.23322849)(6, 0.28589311)(8, 0.34833043)(10, 0.53983274)(12, 0.53222001)(14, 0.35946062000000003)(16, 0.38696663)
    };
    \addlegendentry{Linear}
    
    \addplot[
    color=green,
    mark=halfcircle,
    ]
    coordinates {
    (2, 0.24531943)(4, 0.42656349)(6, 0.70065011)(8, 0.90747243)(10, 1.01678747399999999)(12, 1.58358801)(14, 1.2284076200000001)(16, 1.37575163)
    };
    \addlegendentry{Exact}
\end{axis}
\end{tikzpicture}
\caption{Query time scaling with computing nodes.}
\label{fig:query}
\end{figure*}
We first compare our system to the popular LSH package in PySpark, the current industry standard for ANN queries. We ran the PySpark LSH package on Amazon Web Services using 3-8 m5.xlarge compute nodes, each with 4 cpus and compared the preformance to our system running on an equivalent number of nodes with the same number of CPUs (see figure \ref{fig:spark_comp}). Experiments were done on a subset of the Webspam dataset.

We evaluated the systems based on query time and the average cosine similarity of query vector to the top 128 nearest neighbors returned. We found that the similarity of our system and the PySpark system were nearly identical, but that our system preformed nearly 10,000x faster. Increasing number of computing node does not change the relative scaling. Figure \ref{fig:indexing} shows the indexing time of SLASH with varying cluster sizes on KDD12 and Webspam. We can see the benefits of the model parallelism: the indexing time decreases significantly because the fraction of the dataset on each node decreases. 


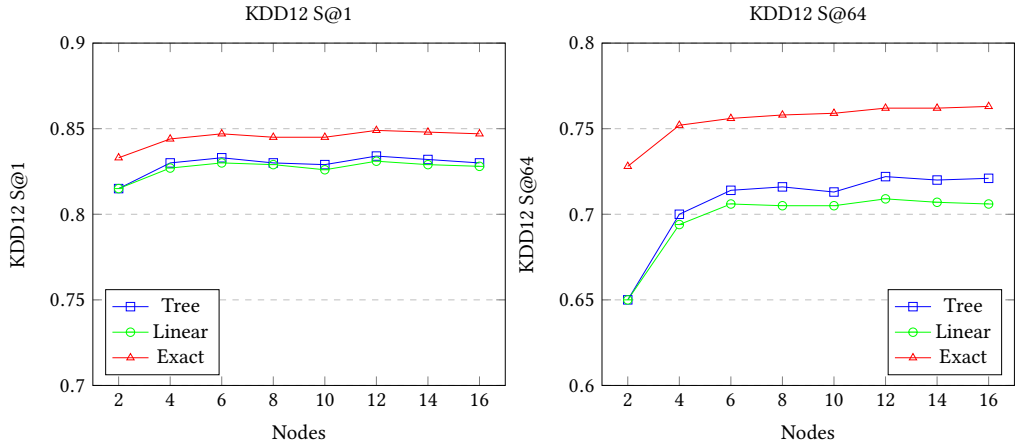
\begin{figure*}
    \centering
\begin{tikzpicture}[scale=0.8]
\begin{axis}[
    title={KDD12 S@1},
    xlabel={Nodes},
    ylabel={KDD12 S@1},
    xmin=1, xmax=17,
    ymin=0.7, ymax=0.9,
    xtick={2,4,6,8,10,12,14,16},
    ytick={0.7,0.75,0.8,0.85,0.9},
    legend pos=south west,
    ymajorgrids=true,
    grid style=dashed,
]
 
\addplot[
    color=blue,
    mark=square,
    ]
    coordinates {
    (2, 0.815)(4, 0.830)(6, 0.833)(8, 0.830)(10, 0.829)(12, 0.834)(14, 0.832)(16, 0.830)
    };
    \addlegendentry{Tree}
  
\addplot[
    color=green,
    mark=halfcircle,
    ]
    coordinates {
    (2, 0.815)(4, 0.827)(6, 0.830)(8, 0.829)(10, 0.826)(12, 0.831)(14, 0.829)(16, 0.828)
    };
    \addlegendentry{Linear}
    
 \addplot[
    color=red,
    mark=triangle,
    ]
    coordinates {
    (2, 0.833)(4, 0.844)(6, 0.847)(8, 0.845)(10, 0.845)(12, 0.849)(14, 0.848)(16, 0.847)
    };
    \addlegendentry{Exact}
\end{axis}
\end{tikzpicture}
\begin{tikzpicture}[scale=0.8]
\begin{axis}[
    title={KDD12 S@64},
    xlabel={Nodes},
    ylabel={KDD12 S@64},
    xmin=1, xmax=17,
    ymin=0.6, ymax=0.8,
    xtick={2,4,6,8,10,12,14,16},
    ytick={0.6,0.65,0.7,0.75,0.8},
    legend pos=south east,
    ymajorgrids=true,
    grid style=dashed,
]
 
\addplot[
    color=blue,
    mark=square,
    ]
    coordinates {
    (2, 0.650)(4, 0.700)(6, 0.714)(8, 0.716)(10, 0.713)(12, 0.722)(14, 0.720)(16, 0.721)
    };
    \addlegendentry{Tree}
  
\addplot[
    color=green,
    mark=halfcircle,
    ]
    coordinates {
    (2, 0.650)(4, 0.694)(6, 0.706)(8, 0.705)(10, 0.705)(12, 0.709)(14, 0.707)(16, 0.706)
    };
    \addlegendentry{Linear}
    
 \addplot[
    color=red,
    mark=triangle,
    ]
    coordinates {
    (2, 0.728)(4, 0.752)(6, 0.756)(8, 0.758)(10, 0.759)(12, 0.762)(14, 0.762)(16, 0.763)
    };
    \addlegendentry{Exact}
\end{axis}
\end{tikzpicture}
\caption{Quality of Retrieval on KDD12 with varying number of machines.}
\label{fig:similarity1}
\end{figure*}

\begin{figure*}
\begin{tikzpicture}[scale=0.8]
\begin{axis}[
    title={Webspam S@1},
    xlabel={Nodes},
    ylabel={Webspam S@1},
    xmin=1, xmax=17,
    ymin=0.8, ymax=1,
    xtick={0,2,4,6,8,10,12,14,16},
    ytick={0.8,0.85,0.9,0.95,1},
    legend pos=south west,
    ymajorgrids=true,
    grid style=dashed,
]
 
\addplot[
    color=blue,
    mark=square,
    ]
    coordinates {
    (2, 0.901)(4, 0.901)(6, 0.901)(8, 0.901)(10, 0.901)(12, 0.901)(14, 0.901)(16, 0.901)
    };
    \addlegendentry{Tree}
  
\addplot[
    color=green,
    mark=halfcircle,
    ]
    coordinates {
    (2, 0.901)(4, 0.901)(6, 0.901)(8, 0.901)(10, 0.900)(12, 0.900)(14, 0.900)(16, 0.900)
    };
    \addlegendentry{Linear}
    
 \addplot[
    color=red,
    mark=triangle,
    ]
    coordinates {
    (2, 0.904)(4, 0.904)(6, 0.904)(8, 0.904)(10, 0.904)(12, 0.904)(14, 0.904)(16, 0.904)
    };
    \addlegendentry{Exact}
\end{axis}
\end{tikzpicture}
\begin{tikzpicture}[scale=0.8]
\begin{axis}[
    title={Webspam S@64},
    xlabel={Nodes},
    ylabel={Webspam S@64},
    xmin=1, xmax=17,
    ymin=0.7, ymax=0.9,
    xtick={0,2,4,6,8,10,12,14,16},
    ytick={0.7,0.75,0.8,0.85,0.9},
    legend pos=south west,
    ymajorgrids=true,
    grid style=dashed,
]
 
\addplot[
    color=blue,
    mark=square,
    ]
    coordinates {
    (2, 0.810)(4, 0.811)(6, 0.811)(8, 0.812)(10, 0.811)(12, 0.812)(14, 0.812)(16, 0.812)
    };
    \addlegendentry{Tree}
  
\addplot[
    color=green,
    mark=halfcircle,
    ]
    coordinates {
    (2, 0.810)(4, 0.810)(6, 0.811)(8, 0.811)(10, 0.811)(12, 0.811)(14, 0.811)(16, 0.811)
    };
    \addlegendentry{Linear}
    
 \addplot[
    color=red,
    mark=triangle,
    ]
    coordinates {
    (2, 0.817)(4, 0.817)(6, 0.817)(8, 0.817)(10, 0.817)(12, 0.817)(14, 0.817)(16, 0.818)
    };
    \addlegendentry{Exact}
\end{axis}
\end{tikzpicture}
\caption{Quality of Retrieval Webspam with varying number of machines.}
\label{fig:similarity2}
\end{figure*}
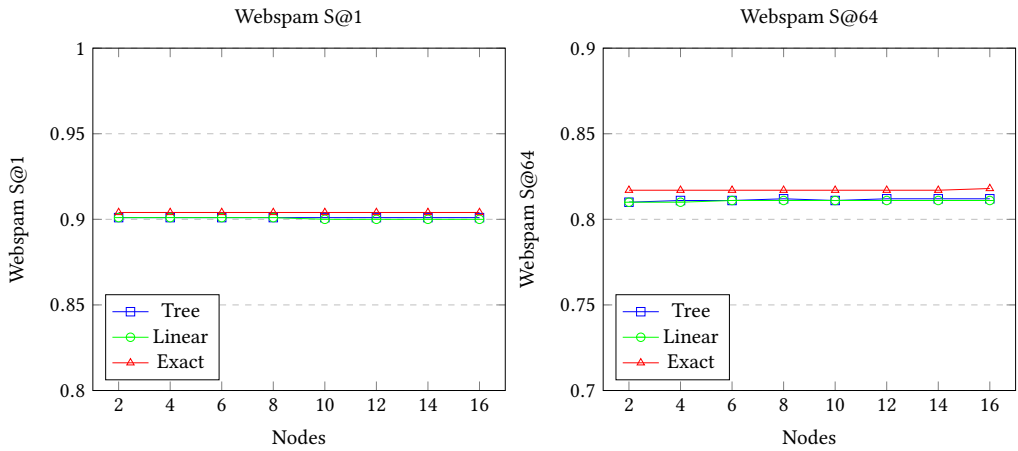
\subsection{Effect of Algorithmic Modifications}

With the two small datasets we rigorously evaluated our different algorithmic modifications: {\bf 1.} Sketching with {\bf linear} aggregation (Linear communication), {\bf 2.} Sketching with smart MPI {\bf tree} aggregation (our proposal), {\bf 3.} standard frequency counting without sketches as done in~\cite{wang}, which we refer to as {\bf exact}. This comparisons are given by Figure \ref{fig:query}.

Finally Figures \ref{fig:similarity1} and \ref{fig:similarity2} show the average cosine similarity of our results on the different datasets using each of the 3 methods. It is important to note that the overall accuracy of our system was not significantly affected by the use of sketches. 

Figure~\ref{fig:query} compares query time.  The use of sketching instead of buckets in~\cite{wang} reduces the memory footprint of communication gaining significantly, even without the tree aggregation. Logarithmic time tree aggregation gives another level of speedup. The query time with increasing number of nodes is remarkably flat, indicating exponential scaling.

\subsection{Indexing and Querying Tera-Scale Criteo Dataset}

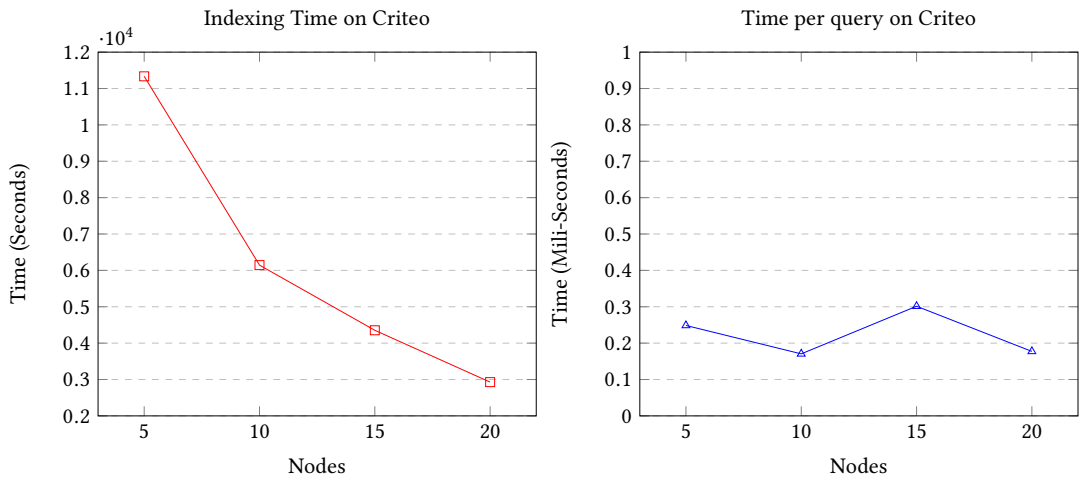
\begin{figure*}
  \centering
\begin{tikzpicture}[scale=0.85]
\begin{axis}[
    title={Indexing Time on Criteo},
    xlabel={Nodes},
    ylabel={Time (Seconds)},
    xmin=3, xmax=22,
    ymin=2000, ymax=12000,
    xtick={5,10,15,20},
    ytick={2000,3000,4000,5000,6000,7000,8000,9000,10000,11000,12000},
    legend pos=north west,
    ymajorgrids=true,
    grid style=dashed,
]
\addplot[
    color=red,
    mark=square,
    ]
    coordinates {
    (5, 11334.8)(10, 6144.75)(15, 4353.4)(20, 2928.29)
    };
\end{axis}
\end{tikzpicture}
\begin{tikzpicture}[scale=0.85]
\begin{axis}[
    title={Time per query on Criteo},
    xlabel={Nodes},
    ylabel={Time (Mili-Seconds)},
    xmin=3, xmax=22,
    ymin=0, ymax=1,
    xtick={5,10,15,20},
    ytick={0,0.1,0.2,0.3,0.4,0.5,0.6,0.7,0.8,0.9,1},
    legend pos=north west,
    ymajorgrids=true,
    grid style=dashed,
]
\addplot[
    color=blue,
    mark=triangle,
    ]
    coordinates {
    (5, 0.248554)(10,  0.170502)(15, 0.301445)(20, 0.177261)
    };
\end{axis}
\end{tikzpicture}
\caption{Indexing and query time on Criteo with computing nodes.}
\label{fig:criteo}
\end{figure*}
For our final experiment we evaluated the scalability of our system on the criteo dataset. We use 12 cores per machine, instead of only 4 cores for other datasets, due to scale. 
This further substantiates our results by showing that as the cluster size doubles, the indexing time is halved, but due to our sketching based queries and aggregation, the query time remains relatively constant (see Figure ~\ref{fig:criteo}). This is primarily because the sketch size only needs to be constant $O(k)$ and aggregating scales logarithmic with the number of nodes with tree aggregation. With 20 nodes, and 12 cores each, we can index 2.3 Terabyte Criteo dataset in a mere 3000 seconds (or about 50 minutes). The query time is around 0.2 millisecond.

\section{Implications and Conclusion}

As the size of today's datasets is quickly eclipsing the size of our computing resources, it is important to acknowledge the ability for intelligent algorithms and data structures to reduce computational overhead and maximize computing resources. While Map-Reduce is a very powerful technique in a variety of applications, if fails to maximize the capabilities of the systems it runs on. By taking a different approach, and combining the speed of LSH with the mergability of a principled sketching algorithm Topkapi, we created a system capable of outperforming the state of the art commercial spark systems by 10,000x. 




\clearpage

\bibliographystyle{ACM-Reference-Format}
\bibliography{sources}

\clearpage

\section{Appendix 1: Proof of Main Theorem}


\subsection{LSH Preliminaries}
Our primary notations and definitions of (c,r)-approximate near neighbor are covered in full rigor Section 1.2 in~\cite{har-peled}. We will simplify them and present them for the sake of completeness. We will use $D(x,y)$ to define the distance.

\begin{definition}
The {\bf $(c,r)$-approximate near-neighbor problem (or $(c,r)-NN)$} with failure probability $f$ is to construct a data structure over the dataset supporting the following query: given any fixed query point $q$, if there exist $x$ in the dataset such that $D(q,x) \le r$, then report some $x'$ such that $D(q,x') \le cr$, with probability $1-f$. 
\end{definition}
We always have $r \ge 0$ and $c \ge 1$. We skip the failure probability $f$ in this or other definitions involving it if it is equal to some absolute constant from $(0,1)$. We can always boost this to any small constant by union bound and paying an extra logarithmic cost. 

\begin{definition}
We call an LSH hash function $(r,cr,p_1,p_2)$-sensitive if for any $x_1$ and $x_2$, we have the following two conditions:
\begin{enumerate}
    \item if $D(x_1,x_2) \le r$ then $Pr(h(x_1) = h(x_2)) \le p_1 $
    \item if $D(x_1,x_2) \ge cr$ then $Pr(h(x_1) = h(x_2)) \ge p_2 $
\end{enumerate}
\end{definition}



Given a query $q$, and $L$ hash tables. In the procedure of Algorithm, the query is mapped to one bucket in each of the hash tables, call it bucket $H(q)$.  Given,  the $(K, L)$ parameterized LSH algorithm, we have the following for given query $q$ and any element $x$ in the dataset:
\begin{enumerate}
    \item if $D(q,x) \le r$, then the probability of finding $x$ in the bucket of the query $H(q)$ (call it $P_x$), in any hash table $i$, is greater than $p_1^K$, i.e. $P_x \ge p_1^K$.
    \item if $D(q,x) \ge cr$, then the probability of finding $x$ in the bucket of the query $H(q)$, in any hash table $i$, is less than $p_2^K$, i.e. $P_x \le p_2^K$.
\end{enumerate}


\begin{figure*}
    \centering
    \includegraphics[scale=0.45]{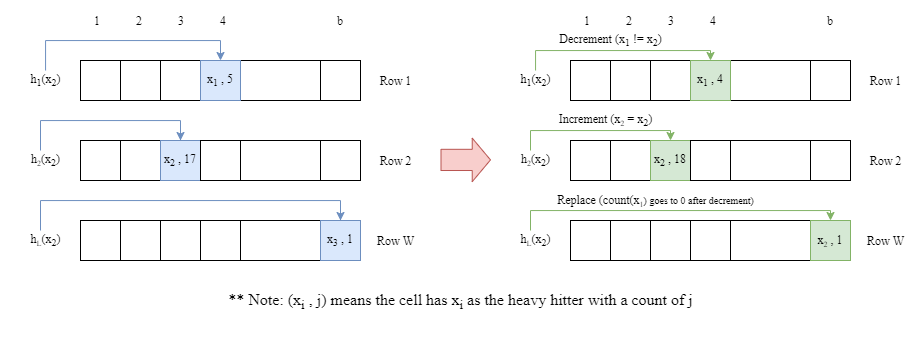}
    \caption{Illustration of Top-K-API Operations}
\end{figure*}
\subsection{Connections to Heavy Hitters on Streams}

Assume that items in each hash buckets of the query form a data stream. We are combining $L$ data streams from each of the $L$ buckets to obtain frequency values of all the elements stream to finally find the most frequent ones. 

Since the hash tables are independent (independent hash functions) and the  data is i.i.d., for every element $x$ in the dataset, we have the frequency of $x$, call it $f_x$, a binomial random variable with parameter $p$ (mean) and $L$ (no of trials). We dont know $p$ but we have conditional bounds. $p \le p_2^K$ if $D(q,x) \ge cr$, when $D(q,x) \le r$ $p \ge p_1^K$.

A simple chernoff bounds tells us that for any $x$, $f_x \in [pL - C \sqrt{L}, pL + C\sqrt{L}]$ with high  probability $1 - f$, where $C > 1$ is some constant. Of course there are factors $\frac{1}{\epsilon^2}$ and $\log{\frac{1}{f}}$ inside the square-root but they are independent of n so does not affect the asymptotics. 

The only thing we need to ensure is that the good neighbors $D(q,x) \le r$ forms the heavy hitters of this stream. For $x$ with $D(q,x) \le r$, we have $f_x \in [p_1^KL - C \sqrt{L}, p_1^KL + C\sqrt{L}]$ with high probability. For sharp concentration of estimate we want (for large enough $C_1$)
$p_1^KL \ge C_1 \sqrt{L}$, which leads to 
\begin{equation}
\label{eq:Kupper}
K \le \frac{\log{\frac{L}{C_1}}}{\log{\frac{1}{p_1}}}
\end{equation}

Alternatively, under proper choice of $K$ and $L$, we can identify the counts of signal (elements $x$ s.t.  $D(q,x) \le r$) from counts of noise (elements $x$ s.t. $D(q,x) \ge cr$) using a heavy hitter algorithm.

In the worst case all the data points ( $n$ of them) are not good neighbors, i.e.,  $D(q,x) \ge cr$ for all of them. In that situation, the noise coming from $D(q,x) \ge cr$ will be of the order $L \times p_1^K \times n$. We want this to be less than the signal $\alpha = L \times p_1^K$, i.e., we want $C_2 \times L \times p_2^K \times n <  L \times p_1^K$. Where $C_2 > 1$ is some constant to ensure enough separation.
\begin{equation}
\label{eq:Klower}
    K \ge \frac{C_2\log{n}}{\log{\frac{p_1}{p_2}}}.
\end{equation}

Thus, with the k-bounded assumption, once all the noise is less than any signal, and we have at most $k$ signals, we have a classical distributed $\Phi$-heavy hitters problem~\cite{mandal}, with $\Phi = \frac{1}{k+1}$, which can be solved with Topkapi of size $O(k)$. 

For feasibility of both conditions (Equation~\ref{eq:Kupper} and~\ref{eq:Klower}), we need 
$$\frac{C_2\log{n}}{\log{\frac{p_1}{p_2}}}\le \frac{\log{\frac{L}{C_1}}}{\log{\frac{1}{p_1}}}$$
or 
$$L \ge C_1n^{\frac{2C_2\log{\frac{1}{p_1}}}{\log{\frac{1}{p_2}} - \log{\frac{1}{p_1}}}}$$
The running time of the algorithms is $O(L)$, so for sub-linearity, we need 
$$\frac{2C_2\log{\frac{1}{p_1}}}{\log{\frac{1}{p_2}} - \log{\frac{1}{p_1}}} < 1 \ \ \ \ \ \text{OR} \ \ \ \ \ \frac{\log{p_1}}{\log{p_2}} \le \frac{1}{2C_2 -1}.$$

$C_2 > 1$, implies  a necessary condition for sub-linearity  to be $\frac{\log{p_1}}{\log{p_2}} \le 0.5$




\section{Appendix 2: Top-K-API Details}

TopkAPI~\cite{mandal} is a simple combination of frequent items~\cite{karp2003simple}  and count sketch~\cite{cormode} based approach which combines the best of the two worlds. It results in a sketching algorithm which is ideal for parallelism and is orders of magnitude faster than best existing implementing of either. It can exploit both local (multi-core) and distributed parallelism ideally suited for MPI implementation. We give a summary of the procedure for brevity. For details on why it works and comparisons with existing sketching approaches see~\cite{mandal}

Top-K-API is a sketch that can be used to estimate the top K most frequent elements of a data stream $S$ with 2 important properties. The first is that its streamable: once the sketch is created you can insert new items without having to dynamically allocate more memory. The second property is that its mergeable: given 2 different sketches you can combine them into a single sketch using a linear pass over the sketches and element wise operations. The new sketch will then provide an estimation of the k most frequent elements in the combined data streams.

\textbf{Initialization:} The data structure is based on a set of $W$ hash functions $\{h_1, h_2,...,h_W\}$, $h_i: \mathbb{R}^d \to [1...W]$, each corresponding to a row of size $B$ in the sketch. The sketch can be updated and queried through the following procedure. For each cell in each row a count is stored along with a heavy hitter, which is a candidate for the top k most frequent elements. At initialization the heavy hitters are null and the counts are 0.

\textbf{Insertion:} To insert an element, $x$ into the sketch foreach row $i$ in the sketch compute the hash index for the row $h_i(x)$. Go to this location in the row and increment the count of the heavy hitter there is $x$, otherwise decrement the counter. If the counter is zero or the heavy hitter is null replace the heavy hitter with $x$ and set the count to 1. See figure 1.

\textbf{Querying:} To obtain a candidate set for the Top K most frequent elements in the data stream, simply select the heavy hitters from the sketch with a corresponding count that is greater than some chosen threshold.

\textbf{Merging:} Finally, 2 top K API sketches can be merged together in a single pass using the following algorithm. Iterate through each corresponding pair of cells in the sketches. If the heavy hitters in the cells match then sum their counts and let the new heavy hitter be the heavy hitter from the cells. If the heavy hitters for the cells differ, then select the new heavy hitter as the one with the larger count and let the new count be the difference of the 2 counts.
\end{document}